\documentclass[superscriptaddress,amsmath,amssymb,aps,pra,twocolumn,floatfix]{revtex4-2}

\usepackage{bm}
\usepackage{graphicx}
\usepackage{float}
\usepackage{placeins}
\usepackage{braket}
\usepackage{bbold}
\usepackage[colorlinks, linkcolor=mycol, citecolor=mycol, urlcolor=mycol, breaklinks]{hyperref}
\usepackage{xcolor}
\usepackage{accents}

\newcommand{\mi}{{\rm i}}
\newcommand{\me}{{\rm e}}
\newcommand{\id}{{\mathbb 1}}

\newcommand{\at}{\text{at}}
\newcommand{\mI}{{I}}
\newcommand{\I}{{\rm Im\,}}
\newcommand{\R}{{\rm Re\,}}
\newcommand{\degree}{^{\circ}}
\definecolor{mycol}{RGB}{10,55,130}

\begin{document}

\title{Collective Excitation Dynamics of a Cold Atom Cloud}
\author{T. S. do Espirito Santo}
\affiliation{Instituto de F\'{i}sica de S\~{a}o Carlos, Universidade de S\~{a}o Paulo - 13566-590 S\~{a}o Carlos, SP, Brazil}
\affiliation{ISIS (UMR 7006) and IPCMS (UMR 7504), Universit\'e de Strasbourg, CNRS, 67000 Strasbourg, France}
\author{P. Weiss}
\affiliation{Universit\'e C\^ote d'Azur, CNRS, Institut de Physique de Nice, France}
\author{A. Cipris}
\affiliation{Universit\'e C\^ote d'Azur, CNRS, Institut de Physique de Nice, France}
\author{R. Kaiser}
\affiliation{Universit\'e C\^ote d'Azur, CNRS, Institut de Physique de Nice, France}
\author{W. Guerin}
\affiliation{Universit\'e C\^ote d'Azur, CNRS, Institut de Physique de Nice, France}
\author{R. Bachelard}
\affiliation{Universit\'e C\^ote d'Azur, CNRS, Institut de Physique de Nice, France}
\affiliation{Departamento de F\'isica, Universidade Federal de S\~{a}o Carlos, Rod.~Washington Lu\'is, km 235 - SP-310, 13565-905 S\~{a}o Carlos, SP, Brazil}
\author{J. Schachenmayer}
\affiliation{ISIS (UMR 7006) and IPCMS (UMR 7504), Universit\'e de Strasbourg, CNRS, 67000 Strasbourg, France}

\begin{abstract}
We study the time-dependent response of a cold atom cloud illuminated by a laser beam immediately after the light is switched on experimentally and theoretically. We show that cooperative effects, which have been previously investigated in  the decay dynamics after the laser is switched off, also give rise to characteristic features in this configuration. In particular, we show that collective Rabi oscillations exhibit a superradiant damping. We first consider an experiment that is performed in the linear-optics regime and well described by a linear coupled-dipole theory. We then show that this linear-optics model breaks down when increasing the saturation parameter, and that the experimental results are then well described by a nonlinear mean-field theory.
\end{abstract}


\maketitle

\section{Introduction}

The optical response of a coherently illuminated cloud of coupled scatterers can dramatically differ from the light emission properties of its individual constituents. Such collective or cooperative effects have been intensively explored in recent years, especially with cold atoms~\cite{Guerin_Light_2017,Kupriyanov2017}. In particular, super- and sub-radiance have been recently investigated in various experimental geometries~\cite{Goban_Super_2015, Guerin_Subra_2016, Araujo_Super_2016, Roof_Obser_2016, Solano:2017, Weiss_Subra_2018, Weiss:2019, Bromley_Colle_2016}. Strikingly,  many recent observations are well explained in the low-excitation limit~\cite{Scully_Direc_2006}, where dynamics can be described by linear equations of motions of classical coupled dipoles~\cite{Svidzinsky_Coope_2010, Courteille_Modif_2010}. It is an important task to explore collective effects beyond this linear-optics regime theoretically~\cite{Arecchi_Atomic_1972, MacGillivray_Theory_1976,Gross_Superradiance_1982,Ott:2013, Zhu_Light_2016, Pucci_Quant_2017, Bettles:2019}, as it is relevant to various contemporary experimental setups studying collective effects with cold atoms~\cite{Westergaard_Observation_2015, Jennewein_Coher_2016, Jennewein_Coher_2018,Georges_Light_2018, Ortiz-Gutierrez_Mollo_2019, Peyrot_Optica_2019, Assemat_Quantum_2019, Muniz_Exploring_2019}.

In recent cold-atom experiments, super- and sub-radiance have been studied by observing the decay dynamics after the driving laser is switched off~\cite{Goban_Super_2015, Guerin_Subra_2016, Araujo_Super_2016, Roof_Obser_2016, Solano:2017, Weiss_Subra_2018, Weiss:2019}.  Here, we demonstrate that the dynamics immediately after the laser switch-on can also be used to observe cooperative effects. We show that in the linear-optics regime the dynamics of the scattered light intensity can be modeled by that of an effective single driven-damped dipole. By fitting a function for the evolution of the intensity emission of this effective dipole~\cite{Jennewein_Coher_2018,guerin_collective_2019},
 we can extract collective decay rates and frequency shifts. The cooperative shifts have been recently understood in terms of a multi-mode collective vacuum Rabi splitting~\cite{guerin_collective_2019}. In this paper we will focus on the collective damping rates and show that they are consistent with those of the experimental observations in the superradiant regime of the switch-off dynamics~\cite{Araujo_Super_2016}.

While most experimental observations are consistent with a linear-optics model in the low-saturation regime, in this paper we also consider the case of larger saturation parameters, and show that experimental signatures start to deviate from the linear model. For this situation, we show that the observed switch-on dynamics can, however, be well described by a non-linear mean-field theory. The agreement between this mean-field model and our experimental data obtained with a dilute atomic sample highlights the importance of high densities for observing quantum effects (beyond the mean-field assumption) in light scattering experiments.

The switch-on scenario studied here provides an alternative approach for studying collective/cooperative effects in cold-atom light-scattering experiments. In contrast to switch-off dynamics, it includes the interplay between coherent excitation and collective dissipation. This increased complexity can give rise to features that may allow to discern quantum-correlations between the atoms more strikingly than in the switch-off scenario, e.g.~in collective Rabi oscillations~\cite{Kaluzny1983}. This can also be advantageous compared to fluorescence measurements, where first- and second-order optical coherences $g^{(1)}$ and $g^{(2)}$ can provide such signatures~\cite{Ott:2013,Pucci_Quant_2017} but require the detection of two-time correlations, using more elaborate techniques such as heterodyne spectroscopy. In a broader context our work analyzes excitations of collective modes in driven-dissipative dynamics of a many-body system. Here we find that in the linear-optics regime, an effective single-mode approximation can remain a decent model. Going beyond the linear and the MF regimes can relate to topics such as driven-dissipative state-preparation of entangled states~\cite{Kraus_Preparation_2008,Muschik_Dissipatively_2011, Zanardi_Coherent_2014} or quantum memories with cold atomic gases~\cite{Dantan_Dynamics_2006, Vasilyev_Quantum_2008,Nunn_Multimode_2008, Afzelius_Multimode_2009}.

This paper is organized as follows: In Sec.~\ref{sec:linear_regime} we present an experiment-theory comparison for the linear-optics regime. We analyze the switch-on dynamics theoretically and show that the experimental data demonstrates superradiance. In Sec.~\ref{sec:beyond} we then proceed to show that for larger saturation parameters the nonlinear mean-field theory provides a better model for the experimentally observed dynamics. Lastly, we conclude and provide an outlook in Sec.~\ref{sec:conclusion}.

\begin{figure*}
\includegraphics[width=\textwidth]{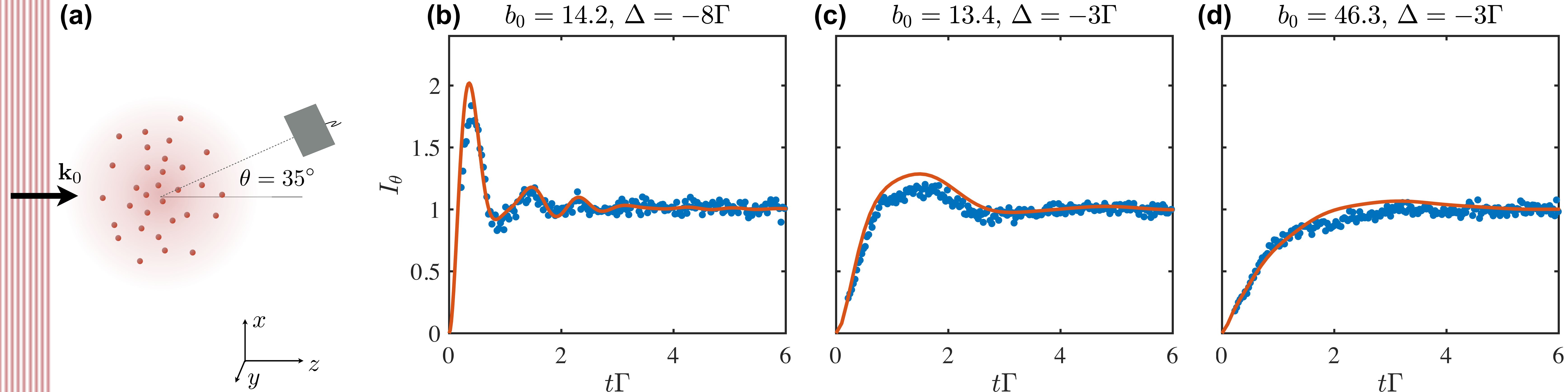}
\caption{(a) Sketch of the experiment: A probe beam (detuned by $\Delta = \omega_L-\omega_0$ from the atomic transition) suddenly illuminates a Gaussian cloud of two-level atoms (transition linewidth $\Gamma$, optical depth $b_0$). The time-dependent scattered light intensity is measured at an angle of $\theta= 35^\circ$ from the axis of the incident beam. (b-d) Experimental time-dependent intensity response for various cloud and laser parameters showing damped Rabi oscillations. The damping depends on the optical depth $b_0$ and $\Delta$. Lines denote a full numerical simulation of linear-optics equations of motion (see text) and reproduce the experiment very well due to the low saturation parameter ($s\approx0.02$).}
\label{fig:fig_intro}
\end{figure*}

\section{Switch-on dynamics in the linear-optics regime}
\label{sec:linear_regime}

In this section, we show how superradiance can be observed in an experiment monitoring the switch-on dynamics of a cold-atom cloud in the linear-optics regime. We start by briefly describing the experimental setup (Sec.~\ref{sec:exp_setup_quick}) and the linear optics model (Sec.~\ref{sec:lin_optics_quick}). We then compare full numerical simulations to experimental data (Sec.~\ref{sec:lin_optics_num_comp}) and show that the cloud dynamics can be modeled by the dynamics of a single effective driven and damped dipole (Sec.~\ref{sec:lin_optics_single_dipole}). We then demonstrate (Sec.~\ref{sec:lin_optics_scaling}) that the damping of the collective oscillation exhibits a superradiant rate, similar to the one observed in the switch-off dynamics~\cite{Araujo_Super_2016}.

\subsection{Experimental setup}
\label{sec:exp_setup_quick}

The experimental data discussed in this section was obtained with the same setup as in~\cite{Araujo_Super_2016}. A precise description of the experiment can thus be found in this reference.

In brief, we produce a three-dimensional Gaussian cloud (rms width $R \approx 1\,\text{mm}$) of $N\approx 10^9$ randomly distributed $^{87}$Rb atoms [see sketch in Fig.~\ref{fig:fig_intro}(a)]. The atoms behave essentially as two-level systems, using the closed atomic $F = 2 \to F' = 3$ transition (wavelength $\lambda = 2\pi c/\omega_0 = 780.24\,\text{nm}$ and linewidth $\Gamma/2\pi = 6.07\,\text{MHz}$). The cloud is homogeneously illuminated by a linearly polarized monochromatic probe beam with beam waist $w\approx5.7\,\text{mm}$, frequency $\omega_L$ and detuning $\Delta = \omega_L - \omega_0$ from the atomic transition. It is propagating along the $z$-direction, $\mathbf{k}_0 = k (0,0,1)^T$, $k = 2\pi/\lambda$. 

Multiple series of pulses with 10\%-90\% rise time of about 6\,ns, which is short compared to the lifetime of the excited state $\tau_\at = \Gamma^{-1} = 26.2\,$ns, are produced by acousto- and electro-optical modulators used in series. During a cycle of pulses the atomic cloud expands balistically, which allows us to probe different on-resonance peak optical depths. The optical depth is defined as $b_0 = \sigma_\textrm{sc}\int \rho(0,0,z)dz$, with $\sigma_\textrm{sc}$ the atomic cross-section. Accounting for the internal structure of Rubidium, this corresponds to $b_0= (7/15) 3N/(k R)^2$ in the experiment. In the scalar-light model used below, the optical thickness is given by $b_0^{S}=2N/(k R)^2$. We also vary the detuning of the probe pulses but then adjust the light intensity accordingly to keep a constant saturation parameter of $s \simeq (2.2 \pm 0.6) \times 10^{-2}$. The time-dependent scattered light intensity is recorded by a photon detector in the far field at an angle of $\theta = 35\degree$ from the $z$-axis.

To clean the recorded intensity signal  from remaining technical imperfections of the light switch-on dynamics, such as small overshoots, we divide the normalized temporal signal recorded with the atoms by a normalized reference profile of the laser without atoms (and white paper as scattering medium). The experimental signal is averaged over a large number of realizations ($\approx 5 \times 10^5$ cycles), and normalized to one in the steady state.

\subsection{Linear-optics dynamics}
\label{sec:lin_optics_quick}

In the limit of low excitation in the cloud, i.e.~in a regime where the atoms are only virtually excited and all population remains in the ground state (see Sec.~\ref{sec:beyond} for a detailed derivation), the dynamics is governed by the well-known linear coupled-dipole (CD) equation
\begin{align}
\frac{d}{dt}\mathbf{b}(t) = \mathbf{M} \mathbf{b}(t) + \mathbf{w}.
\label{eq:cdeqs}
\end{align}
Here, the system is described by the vector of complex excitation amplitudes, $\mathbf{b}(t)$ $=$ $[\beta_1(t),$ $\beta_2(t),$ $\dots,\beta_N(t)]^T$. The laser excitation is governed by the Rabi-frequency vector $\mathbf{w}$ $=$ $-\mi [\Omega_1,$ $\Omega_2,$ $\dots, \Omega_N]^T/2$ , where the complex $\Omega_n = \Omega_0 \me^{\mi \mathbf{k}_0 \cdot \mathbf{r}_n}$ contain the single-atom Rabi frequency as well as the laser phase due to the random positions of the atoms, $\mathbf{r}_n$. Explicitly, the elements of the matrix $\mathbf{M}$ are
\begin{align}
M_{nm} = \delta_{nm} \left(\mi \Delta -\frac{\Gamma}{2}\right) + (\delta_{nm}-1) G_{nm}^s.
\end{align}
The diagonal term governs the single-atom dynamics, while the off-diagonal part includes all long-range dipole-dipole couplings between the atoms. In our setup we consider a cloud of low density, with typical separation between two atoms $n$ and $m$, $r_{nm}  = |{\mathbf r}_n - {\mathbf r}_m|  \gg k^{-1}$. In particular, our experimental peak density of $\rho_0 \approx 0.06 \lambda^{-3}$ corresponds to a typical particle separation of $\bar r_{nm} = 2 \rho_0^{-1/3} \sim 30 k ^{-1}$. Given the large distance between neighbouring atoms the physics will be largely dominated by the dipole-dipole far-field terms. Such a regime is well described by a scalar-light model, for which one finds \cite{Friedberg_Analy_2010}
\begin{align}
G^s_{nm} = \frac{\Gamma}{2}  \frac{\me^{\mi kr_{nm}}}{\mi kr_{nm}}. \label{eq:kernel_scalar}
\end{align}

The formal time-dependent solution of the problem when initially all atoms are in the ground state is
\begin{align}
\mathbf{b}(t) = \left[\me^{\mathbf{M}t} - \id \right]\mathbf{M}^{-1} \mathbf{w}.
\end{align}
The experimentally measured intensity is related to the square of the electric far field at the detector position. Defining the observation direction $\hat{n} = (\sin \theta \cos \phi, \sin \theta \sin \phi, \cos \theta)$, with angles defined for spherical coordinates with the incident laser wave vector $\mathbf{k}_0$ along the $z$ axis, the intensity signal is proportional to
\begin{align}
\mI(\theta, \phi) = \sum_{n,m}^N  \beta_n^* \beta_m \me^{\mi k \hat{n} \cdot (\mathbf{r}_n - \mathbf{r}_m)}.
\end{align}
In our numerical simulations we integrate the signal over the azimuthal angle $\phi$, $I_\theta \equiv \int_0^{2\pi} d\phi\, I(\theta,\phi)$, to reduce small-$N$ fluctuations, and use $\theta = 35\degree$ as in the experimental setup. We always consider the normalized steady-state value $I_{\theta}(t \to \infty) \equiv I_s = 1$.

\subsection{Numerical comparison}
\label{sec:lin_optics_num_comp}

In Fig.~\ref{fig:fig_intro}(b-d) we compare the experimentally recorded time-dependent response signal to a full numerical simulation of the cloud.  Since treating $N=10^9$ atoms is out of reach numerically, even for the linear regime, we model the experiment with an effective cloud consisting of $N_{\rm eff} = 5000$ atoms and a rms radius chosen to match the experimental optical thickness, i.e.~$R = k^{-1}\sqrt{2 N_{\rm eff} / b_0} $. For the results in Fig.~\ref{fig:fig_intro} we take the average over $100$ realizations of random positions of the atoms. For numerical stability we exclude realizations with two atoms much closer than the typical distance between neighbors, here we chose $d_{\rm min} = 0.1 \rho_0^{-1/3}$ with $\rho_0$ the peak density of the effective cloud.

The three panels correspond to different experimental situations with varying detuning and optical thickness: (b) $b_0 = 14.2$ and $\Delta=-8\Gamma$; (c) $b_0 = 13.4$ and $\Delta=-3\Gamma$; and (d) $b_0 = 46.3$ and $\Delta=-3\Gamma$. The intensity evolution is reproduced by the simulations of the linear model very well, although the number of particles used is different from that of the experiments by orders of magnitude. This highlights the central role of the resonant optical thickness as control parameter of the collective dynamics of the dipoles. In order to validate the simulations we checked that results for different $N_{\rm eff}$ (and thus different densities) as well as different $d_{\rm min}$ are indistinguishable from each other. We find the main difference to the experiment in the height of the first oscillation, which is typically lower in the experiment than in the simulation. We attribute this more damped behavior mainly to the finite switch-on time for the laser.

The oscillations after switch-on are generally more damped with increasing $b_0$ and decreasing $|\Delta|$. This is e.g.~seen by the decreased amount of visible oscillations when decreasing $|\Delta|$ at constant $b_0$ [going from panel (b) to panel (c)], and by the even increased damping when keeping $\Delta$ constant and increasing $b_0$ [going from panel (c) to panel (d)]. A systematic study of the damping as function of $\Delta$ and $b_0$ is discussed in Fig.~\ref{fig:fig_gam_scale} below.

Note that the experimental observation does not only excellently agree with the CD simulations, but also with a linear-dispersion theory~\cite{Kuraptsev2017}. This was demonstrated in \cite{guerin_collective_2019}, where we showed that a modification of the oscillation frequency can be very well understood in terms of a multi-mode vacuum Rabi-splitting within this linear-dispersion theory framework.

\subsection{Effective mode}
\label{sec:lin_optics_single_dipole}

\begin{figure*}
\centering
\includegraphics[width=\textwidth]{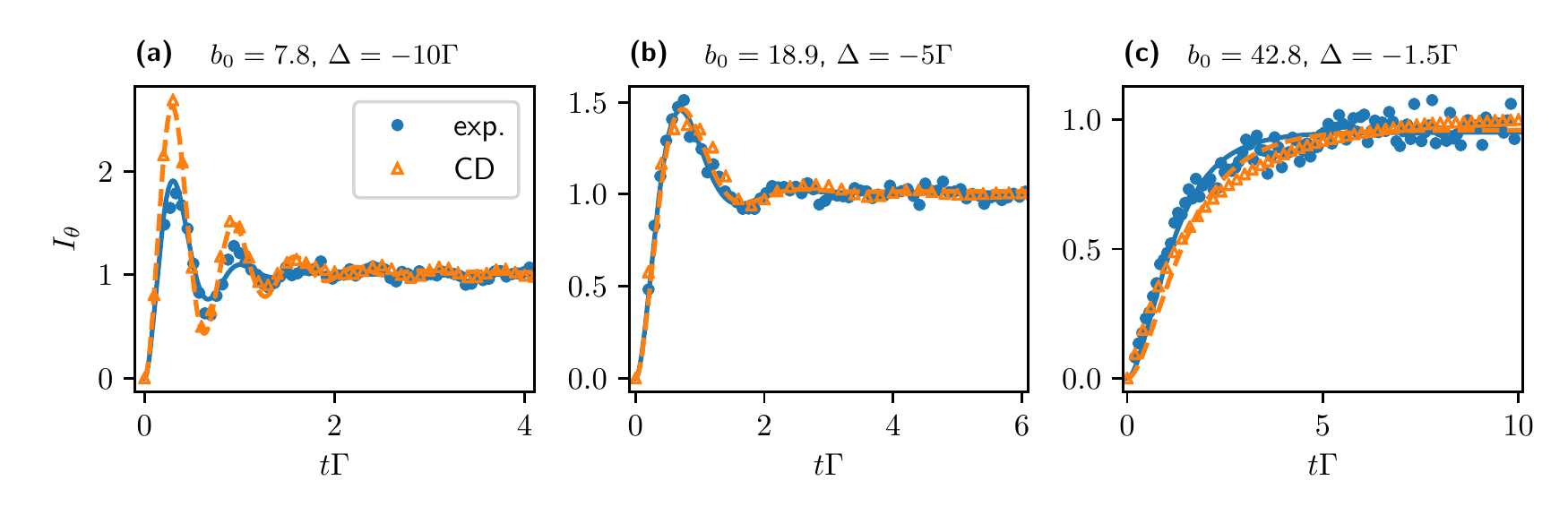}
	\caption{Examples of an effective mode fitting [Eq.~\eqref{eq:single_dipole_fit}] to signals from the experiment (points: data, solid lines: fit) and to CD simulations (triangles: simulation, dashed lines: fit) in various regimes: In the under-damped case (a) the fit to the CD theory is nearly perfect. In the more damped cases (b-c) small deviations from the effective mode (also in CD simulations) are visible.}
	\label{fig:fig_fits}
\end{figure*}

Remarkably, we find that the intensity dynamics of the cloud (averaged over many experimental runs) approximately resembles the evolution of an effective single driven-damped dipole. This implies that we can fit it well to a phenomenological function of the form~\cite{Jennewein_Coher_2018,guerin_collective_2019}
\begin{align}
   	I_\theta = I_s \left| 1-\me^{(\mi \Omega_N - \Gamma_N/2)t}  \right|^2,
   	\label{eq:single_dipole_fit}
\end{align}
where $\Gamma_N$ and $\Omega_N$ denote the decay rate and generalized Rabi frequency of the effective mode, respectively.  Three examples of fits are shown in Fig.~\ref{fig:fig_fits}(a-c) where both the experimental data, and the full numerical simulations have been fitted using Eq.~\eqref{eq:single_dipole_fit}. The fit to CD simulations becomes nearly perfect in the limit of low $b_0$ and large $|\Delta|$ [Fig.~\ref{fig:fig_fits}(a)]. For larger $b_0$ and smaller $|\Delta|$ the fits are still remarkably good and  allow us to extract effective mode properties of the cloud. Similarly, we find that most of the experimental data can be very well fitted by Eq.~\eqref{eq:single_dipole_fit} as also shown in Fig.~\ref{fig:fig_fits}. In the following we discuss the effective mode picture in different limits.

\subsubsection{Single-mode limit (timed Dicke regime)}

The agreement of the single-mode fits to the linear coupled-dipole theory in the limit $|\Delta| \gg \Gamma$ can be understood from the fact that the equilibration dynamics (on the scale of $t\Gamma \sim 10$) is dominated by a single macroscopic scattering mode, which has been e.g.~understood in terms of the excitation of a timed Dicke state~\cite{Scully_Direc_2006, Bienaime_Coope_2013}. Since the dipoles' excitation is essentially determined by the laser in the regime of low optical thickness ($b_0\sim0.1-1$), it is convenient to move to the laser frame by defining
\begin{align}
\widetilde{\beta}_n  = \me^{- \mi \mathbf{k}_0 \cdot\mathbf{r}_n}\beta_n.
\end{align}
This leads to coupled dipole equations of the form
\begin{align}
\frac{d}{dt}\widetilde{\mathbf{b}}(t) = \widetilde{\mathbf{M}} \widetilde{\mathbf{b}}(t) + \widetilde{\mathbf{w}},
\label{eq:cdeqs_laser_frame}
\end{align}
where the Rabi frequency vector is now homogeneous in phase, $\widetilde{ \mathbf{w}} = -\mi (\Omega_0/2) [1,1,\dots,1]^T$, and the coupling matrix $\mathbf{M}$ becomes
\begin{align}
\widetilde{M}_{nm} = \delta_{nm} \left(\mi \Delta -\frac{\Gamma}{2}\right) + (\delta_{nm}-1) G_{nm}^s \me^{\mi \mathbf{k}_0 \cdot (\mathbf{r}_m-\mathbf{r}_n)}.
\label{eq:cd_intmat_laser_frams}
\end{align}

Taking advantage of this uniform driving in the laser frame, we may replace the amplitude vector with identical amplitudes, $\widetilde{\mathbf{b}}(t) \equiv \bar \beta [1,1,\dots,1]^T$. This structure is enforced by taking $\bar \beta$ to evolve as the mean value of all individual coherences, i.e., by summing Eq.~\eqref{eq:cdeqs_laser_frame} over the atoms (this implies that the sum of dipole-dipole interactions for each atom is approximated by its mean value), so one obtains
\begin{align}
    \frac{d}{dt}{\bar{\beta}}(t) =  \left[ \mi \Delta  -\frac{\Gamma}{2} - C \right] \bar \beta(t) -\mi \frac{\Omega_0}{2} ,
    \label{eq:cd_mean_evolve}
\end{align}
with the following complex correction:
\begin{align}
    C 
    &=  \frac{1}{N} \sum_{n,m; n\neq m}^N  G_{nm}^s \me^{\mi \mathbf{k}_0 \cdot (\mathbf{r}_m-\mathbf{r}_n)} .
    \label{eq:C}
\end{align}
By definition, in this approximation the solution to \eqref{eq:cd_mean_evolve} is given by that of a single dipole,
\begin{align}
\bar \beta(t) &= 
\left[ 1 - \me^{
(\mi \Omega_N^{\rm sd} - \Gamma_N^{\rm sd}/{2} ) t}\right] \frac{\mi\Omega_0}{2\mi \Omega_N^{\rm sd} - \Gamma_N^{\rm sd}} ,
\label{eq:single_mode_evolution}
\end{align}
with modified frequency and damping
\begin{align}
\Gamma_N^{\rm sd} &\equiv {\Gamma + 2 \R C} ,\\
\Omega_N^{\rm sd} &\equiv \Delta - \I C .
\end{align}

For the time-dependent intensity signal, this implies
\begin{align}
\mI(t) = |\bar \beta(t)|^2 \sum_{n,m}^N  \me^{\mi (k \hat{n} - \mathbf{k}_0) \cdot (\mathbf{r}_n - \mathbf{r}_m)}.\label{eq:Ids}
\end{align}
In this single-mode limit, the intensity evolution is thus reproduced by a fitting function of the form of Eq.~\eqref{eq:single_dipole_fit}. The geometrical factor in Eq.~\eqref{eq:Ids}, depending on particle and detector position, only modifies the normalization factor, i.e.~the steady-state intensity. The time-dependence of the intensity evolution is independent from the measurement direction in the effective-mode approximation. Note that the geometrical factor features the characteristic $\propto N^2$ enhancement for an intensity measurement in the laser direction, when all terms in the double sum in Eq.~\eqref{eq:Ids} contribute coherently~\cite{Bromley_Colle_2016}. Off-axis measurements only lead to an intensity $\propto N$, since the different random phases for $n\neq m$ in the sum average to zero. Note that this property is generally not found to be true when experimentally studying the collective frequency, where a more elaborate theoretical analysis is necessary~\cite{guerin_collective_2019}.

To compute the value of $C$, we assume a Gaussian distribution of atoms and replace the sums in Eq.~\eqref{eq:C} by an integral. The integration gives
\begin{align}
C  
&= \frac{\Gamma N}{2 (2 \pi)^3 R^6 \mi} \int d\mathbf{r} \int d\mathbf{r^\prime} \frac{\me^{\mi k | \mathbf{r}-\mathbf{r^\prime}|}}{ k |\mathbf{r}-\mathbf{r^\prime}|} \me^{\mi \mathbf{k}_0\cdot(\mathbf{r}-\mathbf{r^\prime})} \me^{\frac{-r^2 + {r^\prime}^2}{2 R^2}}\nonumber\\
&= -\frac{\Gamma}{2} \frac{b_0^{S}}{8} \left[ 2 \mi \frac{D(2 k R)}{\sqrt{\pi}} - (1 - \me^{-4 k^2 R^2}) \right]
\label{eq:C_result}
\end{align}
Here, $D(\dots)$ denotes the Dawson integral, which asymptotically behaves as $D(x \to \infty) \sim 1/(2 x)$. This leads to a shift in frequency of the single mode in dilute clouds that scales with the density, reminiscent of cooperative Lamb shifts~\cite{Friedberg_Frequ_1973}. In our dilute sample ($kR \sim 10^4$ for the experiment and $kR \gtrsim 10 $ for the simulations), the imaginary part of $C$ is much smaller than the transition natural linewidth. Consequently, in our dilute clouds this density-dependent shift  cannot be seen. In contrast, optical thickness-dependent shifts in the oscillation frequencies observed at the switch-on can be interpreted as a multi-mode vacuum Rabi splitting, and represent a measure of the coupling strength of the light modes to the atomic cloud, as shown in~\cite{guerin_collective_2019}. Those shifts are not included in the simplistic single-mode limit. 

As for the decay rate, the exponential term in Eq.~\eqref{eq:C_result} is  strongly suppressed in large clouds, so that
\begin{align}
    C = \frac{\Gamma}{2} \frac{b_0^{S}}{8}.
\end{align}
This implies that in the single-mode limit the damping rate of the effective dipole would be expected to be
\begin{align}
    \Gamma_N^{\rm sd} = {\Gamma} \left(1 + \frac{b_0^{S}}{8}\right).
    \label{eq:single_mode_gamma}
\end{align}
Note that the single-mode assumption thus re-produces the well-known result for the scaling of the decay rate due to collective single-photon superradiance after an excitation of a timed Dicke state \cite{Scully_Direc_2006, Bienaime_Coope_2013}. This can be expected since by assuming an optically dilute cloud, we consider the excitation of the cloud to remain driven mainly by the laser, thus leading to the well-known superradiant behavior of the cloud acting as a single large dipole. Here, we have re-derived this result for the switch-on evolution of the intensity from the cloud. Finally, we note that differently from clouds with homogeneous densities, which may exhibit strong topological effects due to the sharp transition in refractive index at the boundaries (e.g., whispering gallery modes) \cite{Bachelard_Mie_2012, Schilder_Polaritonic_2016}, the smooth density of the Gaussian distributions under consideration prevents a strong contribution from boundary conditions. 

Below, in Fig.~\ref{fig:fig_gam_scale}(a), we demonstrate that the damping parameters obtained from fits to CD simulations follow the predictions from a single-dipole limit only in the limit of very small ``actual'' optical thickness $b(\Delta) \ll 1$, where 
\begin{align}
    b(\Delta) = \frac{b_0}{1+ 4 (\Delta/\Gamma)^2}.
    \label{eq:actual_od}
\end{align}
Note that generally the validity of the timed Dicke state approximation also requires negligible dephasing of the probe-beam across the sample, and therefore $b(\Delta) (\Delta/\Gamma) \ll 1$. In our setup, we expect to be outside of this regime and that multiple modes will be involved in the excitation dynamics.

\subsubsection{Multiple modes}

 To analyze the mode structure it is convenient to diagonalize the coupled-dipole matrix $\widetilde{\mathbf{M}}$ in Eq.~\eqref{eq:cdeqs_laser_frame}. This symmetric complex ({\em not} Hermitian) matrix can always be diagonalized by an orthogonal transformation $\mathbf{A } \widetilde{\mathbf{M}}\mathbf{A}^T = \mathbf{D}$ with $\mathbf{A }^T\mathbf{A } = \id $. Here, $\mathbf{D}$ is a diagonal matrix containing the complex mode eigenvalues $\lambda_\eta$. In the transformed frame, the amplitudes of each mode $\alpha_\eta \equiv (\mathbf{A}\widetilde{\mathbf{b}})_\eta = \sum_m A_{\eta,m} \widetilde{\beta}_m$ evolve as
\begin{align}
{\alpha_\eta}(t) &= \frac{\widetilde w_\eta}{\lambda_\eta} \left( \mathrm{e}^{\lambda_\eta t}   - 1\right).
\label{eq:CD_mode_evolution}
\end{align}
The solution only depends on the complex eigenvalues, and the overlap of the uniform Rabi-frequency vector with the eigenmodes,
$
\widetilde w_\eta = (\mathbf{A}  \mathbf{w})_\eta = \mi \Omega \sum_m A_{\eta,m}
$.
The real and imaginary parts of each eigenvalue, $\lambda_\eta \equiv -\Gamma_\eta/2 -\mi \Omega_\eta $, give rise to a damping and oscillation of the respective mode. As analyzed in \cite{Guerin_Popul_2017} the mode population in the steady state, 
$
|{\alpha_\eta}(t\to \infty)|^2 = {|\widetilde w_\eta|^2}/{|\lambda_\eta|^2}$ 
and depends on the geometrical factor, $|\widetilde w_\eta|^2$, and the spectral factor $1/|\lambda_\eta|^2$. Furthermore, here it becomes evident that the respective mode occupations also depend on time.  It is interesting to note that at short times, at leading order (valid for $t|\lambda_\eta| \lesssim 1$),
$
|{\alpha_\eta}(t \to 0)|^2 = {|\widetilde w_\eta|^2} t^2
$. This implies that for short times the population of the collective modes of the problem only depends on the geometrical factor. The independence from the spectral factor can be understood by the large frequency broadening of the driving laser at the switch-on. This means that in an experiment, the duration of the excitation pulse could be used to control the occupation of the different modes~\cite{Kuraptsev2017}.

The time-dependent intensity signal, for the multi-mode case, can be written in the general form
\begin{align}
\mI(t) 
&= \sum_{\eta,\mu} \alpha_\eta(t) \alpha_\mu(t) \mathcal{G}_{\eta, \mu} \\
\mathcal{G}_{\eta, \mu} &\equiv \sum_{n,m} 
A_{\eta, n}^* A_{\mu, m}  \me^{\mi (k \hat{n} - \mathbf{k}_0) \cdot (\mathbf{r}_n - \mathbf{r}_m)}.
\end{align}
Here, from the geometrical contribution $\mathcal{G}_{\eta, \mu}$ it becomes clear that if the measurement is in the forward direction (coherent scattering), cross-terms between modes play an important role, whereas if the measurement is off-axis and if an angle/realization average is considered (diffuse scattering), different phases from different atoms average to zero, and the dominating contribution to the intensity signal stems from the diagonal mode populations $|\alpha_\eta|^2$.

\medskip

\begin{figure}[b!]
\includegraphics[width=\columnwidth]{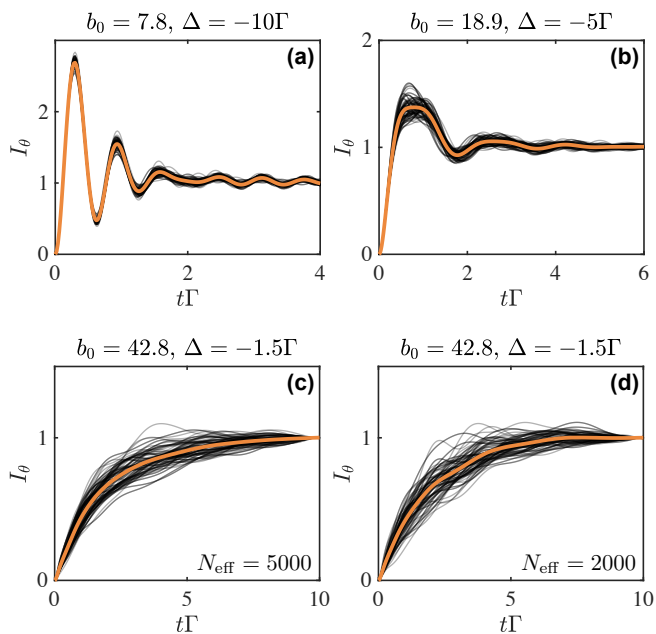}
	\caption{CD simulations for the parameters from Fig.~\ref{fig:fig_fits}. Besides the realization averaged curves (orange lines), we also show 100 single realization results (thin grey lines). (a-c) $N_{\rm eff} = 5000$ (d) $N_{\rm eff} = 2000$. For small $|\Delta|$ and large $b_0$ we observe large fluctuations that are independent on $N_{\rm eff}$.}
	\label{fig:fig_realz}
\end{figure}

We find numerically that considering single atom position realizations, the multi-mode structure can become clearly visible in the intensity signal. This is shown in Fig.~\ref{fig:fig_realz}. There, for CD simulations, besides the averaged intensity signal from Fig.~\ref{fig:fig_fits}, we also show the intensity dynamics for 100 different realizations. For small $b(\Delta)\approx 0.02$ ($b_0 = 7.8$, $|\Delta| = 10$) [Fig.~\ref{fig:fig_realz}(a)] there are only small differences between position realizations, especially for short times, and all realizations follow closely the same curve that can be well fitted by Eq.~\eqref{eq:single_dipole_fit} [Fig.~\ref{fig:fig_fits}(a)]. In contrast, for larger $b(\Delta)$, each realization exhibits very different dynamics already at short times [Fig.~\ref{fig:fig_realz}(b-c) with $b(\Delta) \approx 0.2$ and $b(\Delta) \approx 4.3$, respectively]. Furthermore, those large fluctuations are also robust to the different $N_{\rm eff}$ that are accessible in simulations [compare panels (c) and (d)]. We interpret this behavior as a signature of multi-mode excitations, whose population and structure fluctuate from one realization to another.

Remarkably, we find that the realization averaged signal can still be decently modeled by just the effective mode fitting function~\eqref{eq:single_dipole_fit}, although multiple modes are excited. For example, as we have recently shown \cite{guerin_collective_2019}, for large $b_0$ two frequencies stemming from a multi-mode vacuum Rabi splitting play a crucial role in the dynamics, leading to an imperfect single-mode fit. Nevertheless, the fits still allow us to find effective mode properties of the cloud that are discussed in the next section.

\subsection{Observation of superradiant damping}
\label{sec:lin_optics_scaling}

\begin{figure}
\includegraphics[width=\columnwidth]{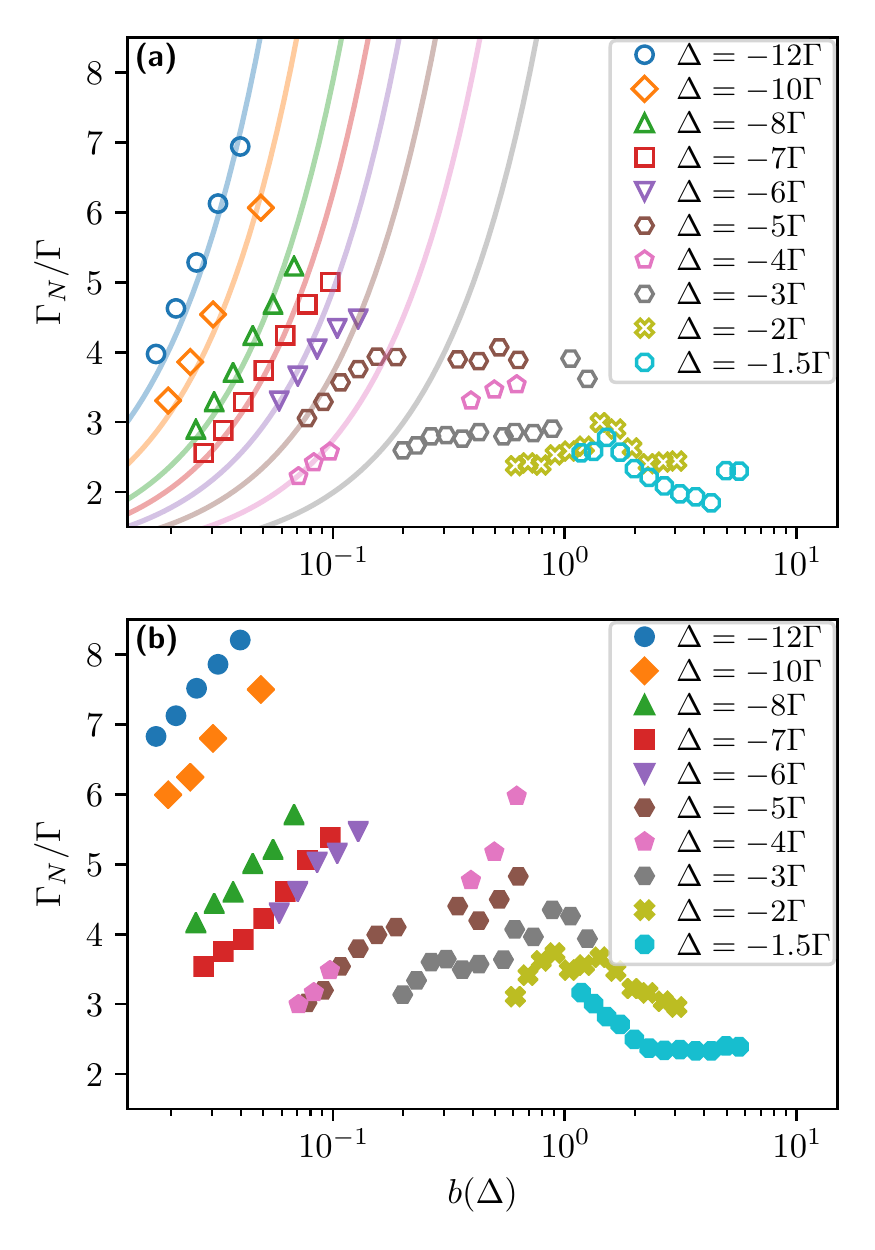}
	\caption{\label{fig:fig_gam_scale} Collective decay rate $\Gamma_N$ showing superradiant behavior, reminiscent of previous switch-off experiments \cite{Araujo_Super_2016}. (a) Fits to CD simulations. (b) Fits to experimental data. We only kept fits for which the fitting function Eq.~\eqref{eq:single_dipole_fit} works well with a value of $R^2>0.85$. For small $b(\Delta)$ the CD theory agrees with the single-mode prediction [lines from Eq.~\eqref{eq:single_mode_gamma}]. Generally, the experiment exhibits a more damped behavior.
	}
\end{figure}

We now analyze the properties of the effective mode of the cold-atom cloud, and show that its excitation dynamics indeed features superradiant decay as previously observed in the switch-off in~\cite{Araujo_Super_2016}. In Fig.~\ref{fig:fig_gam_scale} we summarize the scaling of the damping rate $\Gamma_N$ as a function of the laser detuning and optical thickness for both the CD simulations Fig.~\ref{fig:fig_gam_scale}(a) and the experiment Fig.~\ref{fig:fig_gam_scale}(b). The parameters are extracted from the fitting function~\eqref{eq:single_dipole_fit} of the time-dependent intensity signal. For each fit we evaluate the $R^2$ value, and keep only points with $R^2 > 0.85$.

As shown in Fig.~\ref{fig:fig_gam_scale}, we find that indeed the collective damping shows a superradiant behavior, i.e.~$\Gamma_N>1$. The behavior is very reminiscent of the results for the switch-off dynamics obtained in~\cite{Araujo_Super_2016}, where two regimes of parameters can be identified. When the actual optical thickness is very small, $b(\Delta) \ll 1$, $\Gamma_N$ scales with the resonant optical thickness $b_0$ as it is predicted in the single-mode limit. This regime is highlighted in Fig.~\ref{fig:fig_gam_scale}(a) by the comparison with Eq.~\eqref{eq:single_mode_gamma} (solid lines). In the opposite regime, $b(\Delta) \gtrsim 1$, the superradiant rate presents a reduction due to attenuation and multiple scattering, which can be attributed to an effectively reduced population of superradiant states close to resonance \cite{Guerin_Popul_2017}. Then $b(\Delta)$ becomes the relevant scaling parameter, as seen by a collapse of the points onto a single curve when $\Gamma_N$ is plotted as a function of $b(\Delta)$ \cite{Araujo_Super_2016}. This behavior seems to also appear on the right edge in Fig.~\ref{fig:fig_gam_scale}(a) and (b).

The observed experimental scaling in Fig.~\ref{fig:fig_gam_scale}(b) is similar to the full numerical CD simulation in Fig.~\ref{fig:fig_gam_scale}(a), which are realized using the same $b_0$, yet very different densities and atom numbers. However, especially for the under-damped case $b(\Delta) \lesssim 1$, we find a systematically larger $\Gamma_N$ in the experiment than in the CD simulations. Most relevant dynamics occurs on a very short time scale, and we thus attribute some of this systematic deviation to the finite switch-on time in the experiment. We generally find that a more damped behavior at short times leads to significantly larger estimations of $\Gamma_N$ in the effective mode fit for those data points. Nevertheless, besides a systematic offset to larger values, the $\Gamma_N$ extracted from the experiment exhibit a similar scaling with $b_0$ in the regime of $b(\Delta) \ll 1$.

\section{Beyond the linear-optics regime}
\label{sec:beyond}

While all results in the previous section were for a low saturation parameter in the experiment ($s \approx 0.02$), we now analyze the switch-on dynamics for larger saturation, beyond the linear-optics regime. We first provide a systematic derivation of a mean-field theory (Sec.~\ref{sec:beyond_cd_tools}). Then we show that the mean-field theory is capable of simulating experimentally observed switch-on dynamics for larger saturation (Sec.~\ref{sec:beyond_cd_experiment}).

\subsection{Theory beyond the linear-optics regime} 
\label{sec:beyond_cd_tools}

\subsubsection{Full quantum problem}

Fully quantum mechanically, the system of $N$ two-level atoms is represented by a many-body density matrix $\hat \rho$ which consists of a complex Hermitian $2^N \times 2^N$ matrix. The non-relativistic dynamics of the system is described by a quantum master equation ($\hbar \equiv 1$)~\cite{Stephen_First_1964,Lehmberg_Radia1_1970,Friedberg_Frequ_1973}:
\begin{align}
\frac{d}{dt} \hat \rho = -\mi [\hat H, \hat \rho] + \mathcal{L}(\hat \rho).
\label{eq:meq}
\end{align}
Here, the first part describes coherent Hamiltonian dynamics, i.e., the laser drive and exchange of excitations,
\begin{align}
\hat H &= - \Delta \sum_n  \hat \sigma_n^+  \hat \sigma_n^- +  \frac{1}{2} \sum_n \left( \Omega_n \hat \sigma_i^+  + \Omega_n^* \hat \sigma_n^- \right)  \nonumber \\
&+\sum_{i\neq j} g_{ij} \hat \sigma_i^+ \hat \sigma_j^-, \label{eq:H_dd}
\end{align}
where $\hat \sigma_n^\pm$ denote the standard spin raising and lowering operators for a two-level atom at a position ${\mathbf r}_n$. The system is considered in a frame oscillating at the laser frequency (rotating wave approximation). The complex Rabi frequency corresponds to the one defined in Sec.~\ref{sec:lin_optics_quick} and the coherent coupling is $g_{nm} = \I G^s_{nm}$. Note again that throughout this paper we only consider the scalar-light model as a toy model, and thus neglect the near-field dipole-dipole terms, but they can be easily included by modifying the interaction kernel $G^s_{nm}$~\cite{Friedberg_Analy_2010}.

The second term in Eq.~\eqref{eq:meq} describes dissipative processes in the form of mutual decay, and has the general form
\begin{align}
\mathcal{L}(\hat \rho) &= \sum_{n,m} f_{nm} \left(\hat \sigma_n^- \hat \rho \hat \sigma_m^+ - \{ \hat \sigma_m^+ \hat \sigma_n^-,\hat \rho \}\right) \label{eq:diss}\\
& = \sum_\mu \gamma_\mu\left( 2 L_\mu \hat \rho \hat L_\mu^\dag -\{ \hat L_\mu^\dag \hat L_\mu ,\hat \rho \right) \} .
\end{align}
Here, $\{*,*\}$ denotes the anti-commutator, and the incoherent decay rates are encoded in the symmetric matrix $f_{nm} = \R G_{nm}^s$ (including the $n=m$ elements). In the second line we have written the dissipator in a standard Lindblad form with the jump operators $\hat L_\mu = \sum_n u_{n\mu} \hat \sigma_n^-$ that follow from a diagonalization of the real symmetric matrix  $f_{nm} = \sum_{\mu}  \gamma_\mu u_{n\mu} u_{m\mu}$. The eigenvalues $\gamma_\mu$ determine whether the decay channel is super- or subradiant.

Because of the exponential growth of the Hilbert space with $N$, an exact time-dependent simulation of the full quantum problem is computationally hard and currently limited to $\sim 20$ atoms (using tricks such as quantum trajectories~\cite{Daley_Quant_2014}). The experimental setup is performed in a regime of small density and large optical thickness, for which simulation of much larger systems with $N\sim10^3$ are necessary. In the following we discuss how to reduce the complexity to tackle this regime.

\subsubsection{Mean-field product state ansatz}

\begin{figure*}[tb]
\centering
\includegraphics[width=\textwidth]{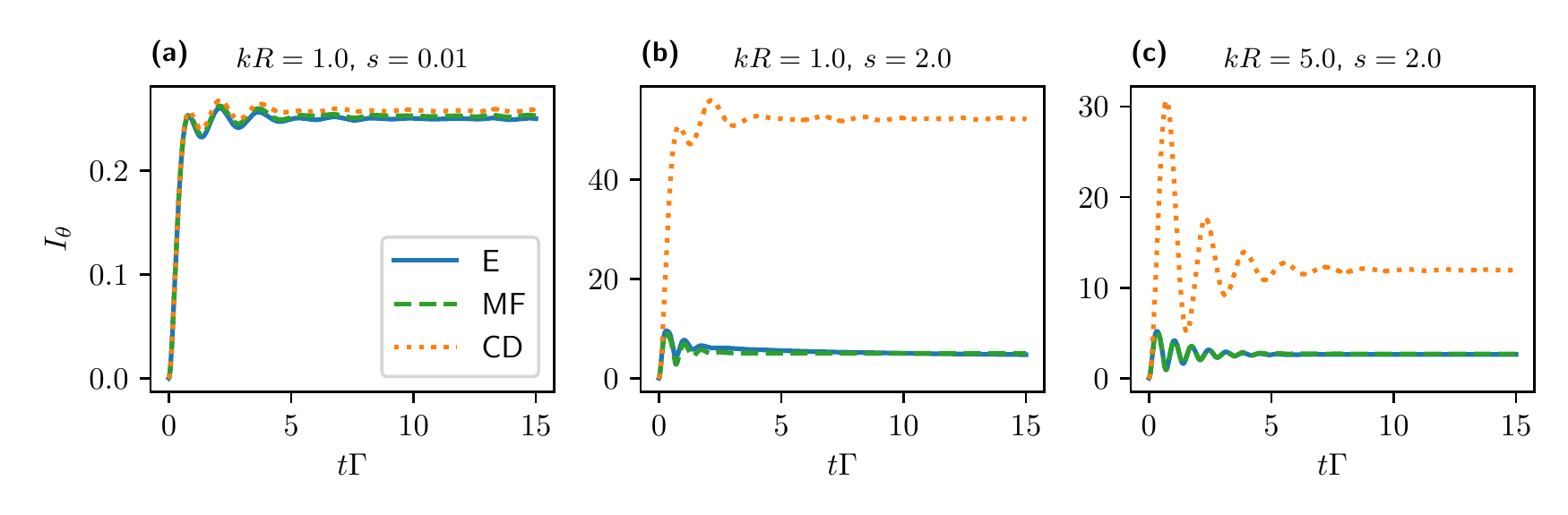}
\caption{Comparison between exact simulations of the master equation (E), Eq.~\eqref{eq:meq}, mean-field  simulations (MF), Eqs.~\eqref{eq:mfb}--\eqref{eq:mfz}, and the coupled-dipole model (CD), Eq.~\eqref{eq:cdeqs}. The intensity is not normalized to the long-time value. We use a small cloud with uniform density ($N_{\rm eff}=6$). Here, $\Delta = -4$ and we use two sphere radii $kR = 1$ (a-b) and $kR = 5$ (c). We tune the Rabi frequency to obtain small and large saturation parameters, $s = 0.01$ (a) and $s = 2$ (b-c), respectively. MF is superior to CD in reproducing the full quantum results for high saturation (c) and only fails for dense clouds {\em and} high saturation (b). Exclusion distance $kd_{\rm min} = 0.1$, average over $21$ realizations.}
\label{fig:fig_bench_mf}
\end{figure*}

To drastically reduce the size of the Hilbert space, a common ansatz is to neglect any type of entanglement between the atoms. In such a mean-field (MF) situation the full density matrix is considered to remain in a product state of the form
\begin{align}
\hat \rho = \prod_n \hat \rho^s_n. \label{eq:fact}
\end{align} 
Enforcing this factorized form at all times leads to time-dependent equations for each local density matrix $\hat \rho^s_n$
\begin{align}
&\frac{d}{dt} \hat \rho_n^s = \mathrm{Tr}_{m\neq n}\left(\frac{d}{dt} \hat \rho \right).
\end{align}
Due to the partial trace operation on the right hand side the MF equations of motion become nonlinear.  Note that the factorization property from~Eq.~\eqref{eq:fact} has also been used in the Maxwell-Bloch description of a high-saturation regime in~\cite{Jennewein_Coher_2018}.

Every single-atom density matrix, $\hat \rho^s_n$, can be parameterized by the complex expectations of $\beta_n = \langle \hat \sigma_n^- \rangle$ and by $z_n = \langle \hat \sigma_n^+ \hat \sigma_n^- - \hat \sigma_n^- \hat \sigma_n^+ \rangle$ via $\hat \rho_n^s = (\id + 2 \beta_n^* \hat \sigma_n^- + 2 \beta_n \hat \sigma_n^+ + z_n \hat \sigma_n^z)/2$.  Then, the Hamiltonian \eqref{eq:H_dd} and dissipator \eqref{eq:diss} lead to the following compact form of the MF equations:
\begin{align}
&\dot \beta_n = \left(\mi \Delta -\frac{\Gamma}{2}\right) \beta_n +\mi W_n z_n, \label{eq:mfb}\\
&\dot z_n = - \Gamma (1 + z_n) - 4 \text{Im}(\beta_n W_n^*). \label{eq:mfz}
\end{align}
Here, we defined the general field acting on atom $n$ as
\begin{align}
& W_n = \frac{\Omega_n}{2} -\mi \sum_{m\neq n}  G^s_{nm} \beta_m.
\end{align}
Importantly the number of the nonlinear set of equations in Eq.~\eqref{eq:mfb} and \eqref{eq:mfz} only scales linearly with the system size, and thus a direct numerical integration is still feasible for thousands of atoms. The physics behind the MF model is quite evident: all the atoms $m$ create a mean field that acts upon dipole $n$, $W_n$. The ``coherent'' real part of $W_n$ drives atom $n$ just like the external laser, it comes from the virtual photon exchange in the dipole-dipole couplings. The ``incoherent'' imaginary part of $W_n$ gives  rise to damping, and also to non-trivial evolution, which can lead to effects such as synchronization~\cite{Xu_Synch_2014,Zhu_Synch_2015}.

\begin{figure*}
\centering
\includegraphics[scale=1]{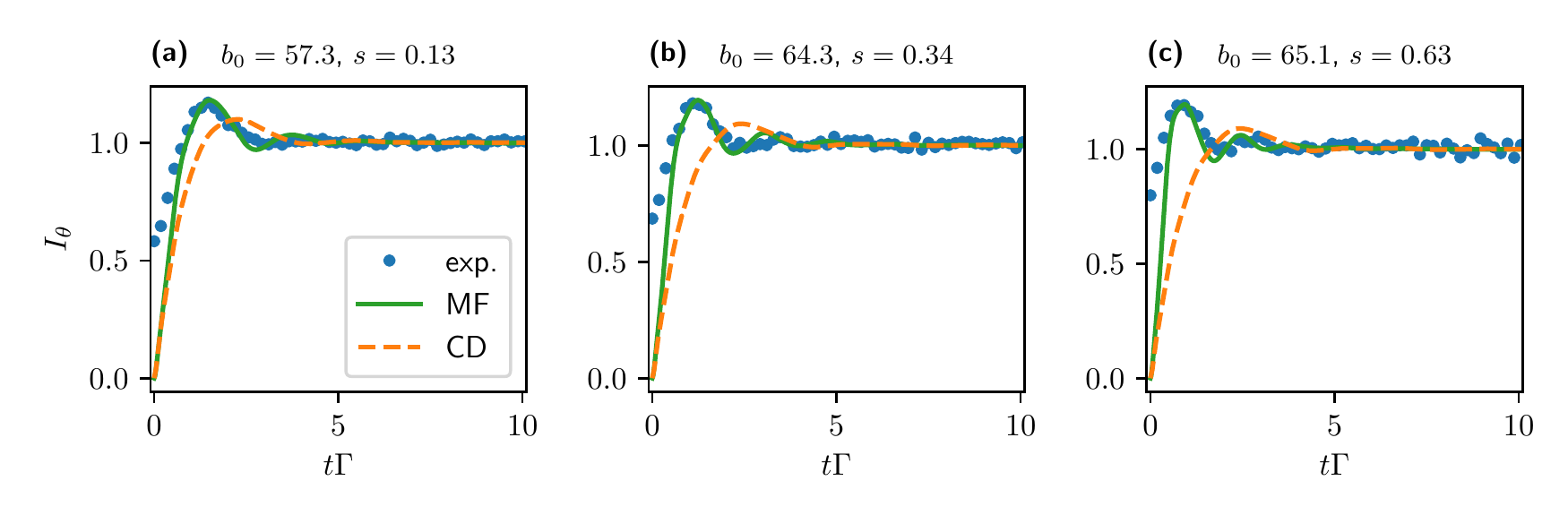}
\caption{Comparison between the normalized intensity evolution from the experiment with MF and CD simulations in a large saturation regime ($s \gtrsim 0.1$) and with $b_0 \sim 60$: (a) $s=0.13$, (b) $s=0.34$, (c) $s=0.63$. Here, $\Delta = -4 \Gamma$. For the simulations $N_{\rm eff}=5000$ and results for $78$ random atom positions have been averaged.} 
\label{fig:fig_bench_exp_highsat}
\end{figure*}

The MF equations~\eqref{eq:mfb} and \eqref{eq:mfz} become linear in the low-excitation limit, as in the full quantum case. Then, one can approximate $z_n\approx-1$ at all times, and recover the coupled-dipole equations
\begin{align}
&\dot \beta_n = \left(\mi \Delta -\frac{\Gamma}{2}\right) \beta_n -\mi W_n, 
\end{align}
which are identical to Eq.~\eqref{eq:cdeqs}.

In Fig.~\ref{fig:fig_bench_mf} we analyze the validity of the MF approximation by studying the switch-on dynamics for a small toy-model cloud consisting of $N_{\rm eff}=6$ atoms. Here, we uniformely distribute the atoms in a sphere with different radii $kR$. For this small system we compare an exact simulation of the master equation [Eq.~\eqref{eq:meq}] to the prediction given by the MF product state ansatz [Eqs.~\eqref{eq:mfb}--\eqref{eq:mfz}] and the coupled-dipole model [Eq.~\eqref{eq:cdeqs}]. Note that for numerical stability, in the toy-model setup of Fig.~\ref{fig:fig_bench_mf} we still only consider far-field terms in the interaction kernel, although at such close distances near-field interaction terms would play the dominant role.

For a small saturation parameter, here defined by $s = {2\Omega_0^2}/{(4 \Delta^2 + \Gamma^2)}$, [$s = 0.01$ in Fig.~\ref{fig:fig_bench_mf}(a)] we find that, as expected, all three models agree relatively well with each other, even though the effective density is very high ($kR=1$, sphere density $\rho_s \sim 1.4 k^{3}$). For a larger saturation of $s = 2.0$ in the high-density sphere, the linear coupled-dipole model provides very inaccurate results as seen in Fig.~\ref{fig:fig_bench_mf}(b). Here, the MF result fails in capturing a slow decreasing slope for the intensity at later times, but still reproduces the frequency of the oscillation reasonably well. For a larger sphere ($kR=5$, sphere density $\rho_s \sim 0.01 k^{3}$) and high saturation of $s=2.0$ we observe that MF can perfectly capture the exact intensity evolution [Fig.~\ref{fig:fig_bench_mf}(c)], while the CD simulation  clearly fails. Note that here we do not normalize the intensity evolution in the long-time limit. The CD simulations do not feature any population inversion and thus lead to very inaccurate predictions for the steady-state magnitude of the intensity for larger saturation parameters in Figs.~\ref{fig:fig_bench_mf}(b) and (c).

The only approximation for our MF simulation is the product-state assumption from Eq.~\eqref{eq:fact}, which limits the amount of possible quantum-correlations between the atoms. For example, entangled pure atomic states are not supported by this ansatz. Our results thus highlight the importance of strong correlations between atoms for observing quantum effects in light-scattering experiments. From our comparisons we conclude that for studying the time-dependent intensity evolution, inter-atomic correlations are mostly relevant in the high density limit. For very closely spaced atoms, large Hamiltonian interactions can induce strong correlations between atoms leading to a breakdown of the MF ansatz. In contrast, for a relatively dilute cloud (for which the optical thickness can still be high) we find that the MF assumption provides valid results. Importantly, in the low density scenario, MF simulations then also provide good estimations for large saturation. Note that one can include more quantum correlations between atoms by further adding e.g.~two-atom correlation observables in the equations of motions~\cite{Pucci_Quant_2017}. Such corrections, however, come at the price of increasing the complexity of simulations to $\sim N^2_\mathrm{eff}$, and currently limit simulations to systems with a few hundreds of atoms.

\subsection{Comparison with the experiment beyond the linear-optics regime} 
\label{sec:beyond_cd_experiment}

Finally, we  compare experimental switch-on dynamics in a high-saturation regime to simulations. The experimental data discussed in this section have been taken on the same apparatus as in Sec.~\ref{sec:linear_regime} with a few upgrades described in Refs.~\cite{Weiss_Subra_2018, Weiss:2019}. However, the 10\% to 90\% rise time of the probe laser is now slightly longer, about 17\,ns, because only acousto-optical modulators are used to produce the pulses. This results in a more diffuse and slow on-set of the intensity signal. In comparisons with theory we compensate for this by shifting the time-signals to match the first peak position. For all data points we find good agreements with a shift of $\sim \Gamma^{-1}$, consistent among the panels in Fig.~\ref{fig:fig_bench_exp_highsat}.

In Fig.~\ref{fig:fig_bench_exp_highsat} we show results, at large saturation parameters, and  compare between experimental data and the mean-field predictions, as well as the linear-optics CD simulations. We show results for a large optical thickness of $b_0 \sim 60$ and for an increasingly large saturation parameter of $s=0.13$, $s=0.34$, and $s=0.63$ in panels Fig.~\ref{fig:fig_bench_exp_highsat}(a-c), respectively. It is striking that while the CD simulations are capable of describing the experiment for a value of $s\approx 0.02$ in Sec.~\ref{sec:linear_regime}, here for $s \gtrsim 0.1$ this linear model is insufficient. Moreover with increasing $s$, we observe that the CD prediction (which does not depend on $s$ due to the linearity of the CD equations) becomes worse. The MF simulations, in contrast, predict the trend of the experimental data of exhibiting a more pronounced oscillation with increasing $s$.

\section{Conclusion \& Outlook}
\label{sec:conclusion}

We have demonstrated that the time evolution of the intensity of laser light scattered off a cold-atom cloud can be used to observe collective effects, in particular superradiance. Here we have shown that superradiance can not only be observed after the laser is rapidly switched off, as in \cite{Araujo_Super_2016}, but also in the damping of oscillations immediately after the laser is switched on.

In a limit of low saturation (low intensity), the dynamics can be described by a linear-optics coupled-dipole model, which matches the experimental behavior very well. For low optical thickness, the results for superradiant damping follow the predictions of a single ``mean'' mode approximation (timed Dicke state excitation). In general, and especially for larger optical thickness, multiple modes are excited. Nevertheless, the cloud can still be reasonably well modeled by an effective single-mode response.  

We furthermore showed that when the saturation is increased, the coupled-dipole model becomes insufficient, as expected. Instead, for this regime we find that experimental data can be well described by a simulation of non-linear mean-field equations that follow from a product state ansatz. We showed that this efficient numerical approach works well for simulating dynamics in dilute clouds with large excitation fractions and large optical thickness.

It will be interesting to analyze signatures going beyond the coupled-dipole and mean-field assumptions discussed here, which has been a topic of recent interest~\cite{Ott:2013, Zhu_Light_2016, Pucci_Quant_2017, Ortiz-Gutierrez_Mollo_2019, Bettles:2019}. Here, in particular we showed that the time-dependent switch-on intensity signal does not discern such signatures unless going to a high-density regime. There the time-dependent response could be a useful tool for quantum signatures \cite{Jennewein_Coher_2018, Jennewein_Coher_2016, Bromley_Colle_2016, Peyrot_Optica_2019, Schilder_Near-r_2019}. Beyond mean-field corrections could be included using approaches that take quantum correlations between pairs of atoms into account~\cite{Pucci_Quant_2017}, by exploiting a semi-classical phase space approach~\cite{Zhu_Agen_2019} or a combination of both~\cite{Pucci_Simul_2016}.

\section{Acknowledgements}
We thank Ivor Kre\v{s}ic and Michelle Ara\'ujo for their contribution in setting up the fast switch-on system and Luis Orozco for fruitful discussions. Part of this work was performed in the framework of the European Training Network ColOpt, which is funded by the European Union (EU) Horizon 2020 programme under the Marie Sklodowska-Curie action, grant agreement No.~721465. R.B.~and T.S.d.E.S. benefited from Grants from S\~ao Paulo Research Foundation (FAPESP) (Grants Nrs.~2017/10294-2, 2018/01447-2, 2018/15554-5 and 2019/02071-9) and from the National Council for Scientific and Technological Development (CNPq) Grant Nrs.~302981/2017-9 and 409946/2018-4. R.B.~and R.K.~received support from project CAPES-COFECUB (Ph879-17/CAPES 88887.130197/2017-01).  P.W.~is supported by the Deutsche Forschungsgemeinschaft (grant WE 6356/1-1). J.S.~is supported by the French National Research Agency (ANR) through the Programme d'Investissement d'Avenir under contract ANR-11-LABX-0058\_NIE within the Investissement d'Avenir program ANR-10-IDEX-0002-02. Research was carried out using computational resources of the Center for Mathematical Sciences Applied to Industry (CeMEAI) funded by FAPESP (grant 2013/07375-0) and the Centre de calcul de l'Universit\'e de Strasbourg.

\bibliographystyle{apsrev4-2}
\bibliography{exc_dyn_bib}{}

\end{document}